\documentclass[%
 aip,
 amsmath,amssymb,
 reprint,%
]{revtex4-1}

\usepackage{graphicx}
\usepackage{dcolumn}
\usepackage{bm}

\usepackage[utf8]{inputenc}
\usepackage[T1]{fontenc}
\usepackage{mathptmx}

\newcommand{\mx}{\text{max}}

\newcommand{\QSS}{\text{QSS}}

\usepackage{url}
\usepackage[version=3]{mhchem}
\usepackage{color}
\usepackage{amsmath}
\usepackage{lipsum}
\usepackage{amssymb}
\usepackage[version=3]{mhchem}
\usepackage{mathrsfs}      
\usepackage{booktabs}   
\usepackage{subfigure}
\usepackage{float}
\usepackage{appendix}
\usepackage{comment}

\usepackage{epsfig}
\usepackage{dcolumn}%
\usepackage{relsize}

\begin{document}

\title[Non-Boltzmann Vibrational Energy Distributions and Coupling to Dissociation Rate]{Non-Boltzmann Vibrational Energy Distributions and Coupling to Dissociation Rate}

\author{Narendra Singh}
 \email{singh455@umn.edu.}
\author{Thomas Schwartzentruber}%
\affiliation{ 
Department of Aerospace Engineering and Mechanics, 
University of Minnesota, Minneapolis, MN 55455, USA}%


\date{\today}

\begin{abstract} In this article, we propose a generalized model for nonequilibrium vibrational energy distribution functions. The model can be used, in place of equilibrium (Boltzmann) distribution functions, when deriving reaction rate constants for high-temperature nonequilibrium flows. The distribution model is derived based on recent \textit{ab initio} calculations, carried out using potential energy surfaces developed using accurate computational quantum chemistry techniques for the purpose of studying air chemistry at high temperatures. Immediately behind a strong shock wave, the vibrational energy distribution is non-Boltzmann. Specifically, as the gas internal energy rapidly excites to a high temperature, overpopulation of the high-energy tail (relative to a corresponding Boltzmann distribution) is observed in \textit{ab initio} simulations. As the gas excites further and begins to dissociate, a depletion of the high-energy tail is observed, during a time-invariant quasi-steady state (QSS). Since the probability of dissociation is exponentially related to the vibrational energy of the dissociating molecule, the overall dissociation rate is sensitive to the populations of these high vibrational energy states. The non-Boltzmann effects captured by the new model either enhance or reduce the dissociation rate relative to that obtained assuming a Boltzmann distribution. This article proposes a simple model that is demonstrated to reproduce these non-Boltzmann effects quantitatively when compared to \textit{ab initio} simulations.
\end{abstract}





\maketitle

\section{Introduction}
The strong shock wave created by a hypersonic vehicle increases air temperature to thousands of degrees. At such temperatures, molecular oxygen and nitrogen dissociate into reactive atomic species which can cause heat shield ablation. In order to predict the state of the gas near the vehicle surface, the heat flux to the surface, and the gas-surface chemistry, the extent of dissociation that occurs between the shock wave and the vehicle surface must be accurately predicted. Under hypersonic conditions, the flow downstream of the shock wave is in a state of strong thermochemical nonequilibrium, and predicting the extent of dissociation is challenging due to its coupling with internal energy excitation\cite{candler2019rate}, which occurs at a finite-rate.
State-of-the-art computational fluid dynamics (CFD) solvers model this coupling by solving an additional energy equation for the gas vibrational energy (some codes also solve a separate equation for rotational energy). The chemical kinetics models used in such CFD codes incorporate local thermochemical nonequilibrium, where equi-partition of energy does not hold, using the `multi-temperature framework' \cite{park1989assessment}. This involves defining and tracking pseudo temperatures such as vibrational temperature $(T_v)$ (and sometimes rotational temperature, $T_{rot}$), where finite-rate energy exchange between average vibrational energy and translational energy is modeled using a modified version of the Landau-Teller equation \cite{landau1936theorie}. Since both trans-rotational energy and vibrational energy are separate quantities that are updated within each computational cell at each simulation timestep, reaction rate constants are then parametrized in terms of both $T$ and $T_v$ to introduce coupling. In practice, the combined models \cite{park1989assessment,park1993review,park1988two,marrone1963chemical,knab1995theory}, including both internal energy relaxation and dissociation are parameterized empirically using a rather limited amount of shock-tube experimental data \cite{byron1966shock,appleton1968shock,hanson1972shock} from the literature. Since the experimental measurements are unable to probe the details of vibration-dissociation coupling, including non-Boltzmann vibrational energy distributions (where $T_v$ may not even be an appropriate model parameter), there is uncertainty that these empirical models are accurate for conditions outside of the limited experimental data.

An alternative approach to the multi-temperature framework is to solve the full-master equation, which includes each quantized rovibrational energy state as an individual species in the calculation. If all bound-bound and bound-free transition rates can be specified, then the evolution of the gas internal energy state and coupling to dissociation reactions can be computed, including non-Boltzmann rovibrational energy distributions. 
In principle, these transition rates can be determined using \textit{ab initio} quasi-classical-trajectory (QCT) \cite{truhlar1979atom} collision calculations. Recently, accurate potential energy surfaces (PESs) have been developed specifically for studying high temperature air, including PESs for N$_2$--N$_2$ and N--N$_2$ collisions  \cite{paukku2013global,paukku2014erratum}, O$_2$--O$_2$ \cite{paukku2017potential} and O--O$_2$ \cite{O2OTruhlar} collisions, as well as N$_2$--O$_2$ collisions \cite{varga2016potential} and N$_2$--O collisions \cite{lin2016global}). The full set of transition rates have been determined via QCT and used in master equation analysis for certain atom-diatom systems; N$_2$+N \cite{macdonald2018_QCT,macdonald2018construction_DMS} and O$_2$+O \cite{andrienko2016rovibrational}. The key physical processes that master-equation analysis can reveal, that multi-temperature models can not, is the evolution of non-Boltzmann internal energy distributions and their state-resolved coupling to dissociation processes. 
However, the vast number of transition rates ($\approx 10^{15}$ transition rates for N$_2$+N$_2$ and $\approx 10^7$ for N$_2$+N) required for full master equation analysis makes the approach computationally intractable for the full air system. To circumvent this issue, coarse-graining (i..e. grouping) of quantized states has been used \cite{macdonald2018_QCT,macdonald2018construction_DMS}. Such approaches involve assumptions and binning-models, and achieving a feasible reduction in state space while maintaining the desired accuracy is still an open problem \cite{macdonald2018_QCT}. 

Instead of master equation analysis, our approach involves the use of Direct Molecular Simulation (DMS); a complete description can be found in Refs. \cite{tomDMS,torres2019direct}. Within DMS, a trajectory calculation is performed (on a PES) for each collision within a direct simulation Monte Carlo (DSMC) calculation \cite{boyd2017nonequilibrium,bird1981monte}. Essentially, instead of precomputing the probabilities of all possible energy transitions through large numbers of QCT calculations, the DMS method performs collision
trajectories ``on the fly'' within a simulation of an evolving gas
system. In this manner, the molecular state resulting from one
collision becomes the initial state for the next collision. Instead of
resolving all possible energy transitions, the DMS method
automatically simulates only the most dominant energy transitions
that actually occur with non-negligible frequency for the conditions of
interest. Although computationally demanding, this approach is now
tractable for diatom-diatom systems \cite{valentini2016dynamics,grover2017internal,grover2019direct} and, in fact, for gas mixtures
with a number of species combinations within collisions.
As a result, DMS can also be viewed as a means of importance sampling for the vast number of state-to-state transitions, since the simulated collisions are those that actually occur within an evolving gas system of interest (i.e. relevant post-shock conditions).
Alternatively, DMS can be viewed as an acceleration technique for the molecular dynamics (MD) simulation of dilute gases, where the PES is the sole model input.

We use DMS results for relevant post-shock conditions to study non-Boltzmann internal energy distribution functions and coupling to dissociation. We use the results to guide the development of a simple model that captures the most important non-Boltzmann processes. Specifically, this article begins by analyzing DMS results for the vibrational energy distributions produced by a range of post-shock conditions, a generalized model is then proposed and verified to accurately reproduce the DMS results. Finally, this article outlines how the new model can be used to analytically derive continuum multi-temperature dissociation rate expressions that include terms accounting for non-Boltzmann internal energy coupling to dissociation. Furthermore, the general form of the nonequilibrium distribution model may be useful for applications outside of hypersonics.

  \begin{figure}
  \subfigure[]{
   \includegraphics[width=3.0in,trim={0.00cm 0.00cm 0.0cm 1.0cm},clip]{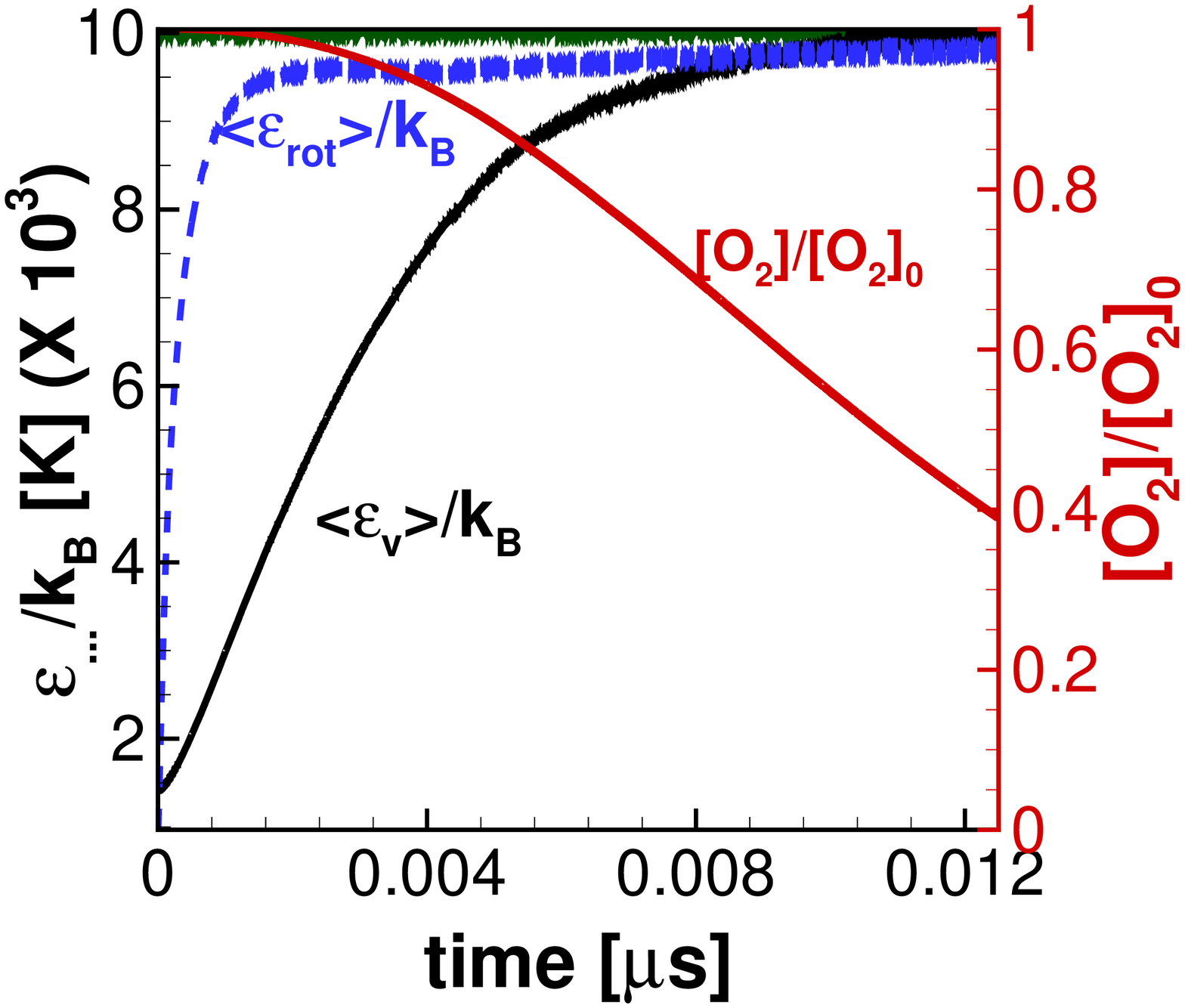}
   \label{10K_excitation_DMS}
  }
\subfigure[]{
\includegraphics[width=3.0in,trim={0.00cm 0.0cm 0.0cm 1.0cm},clip]{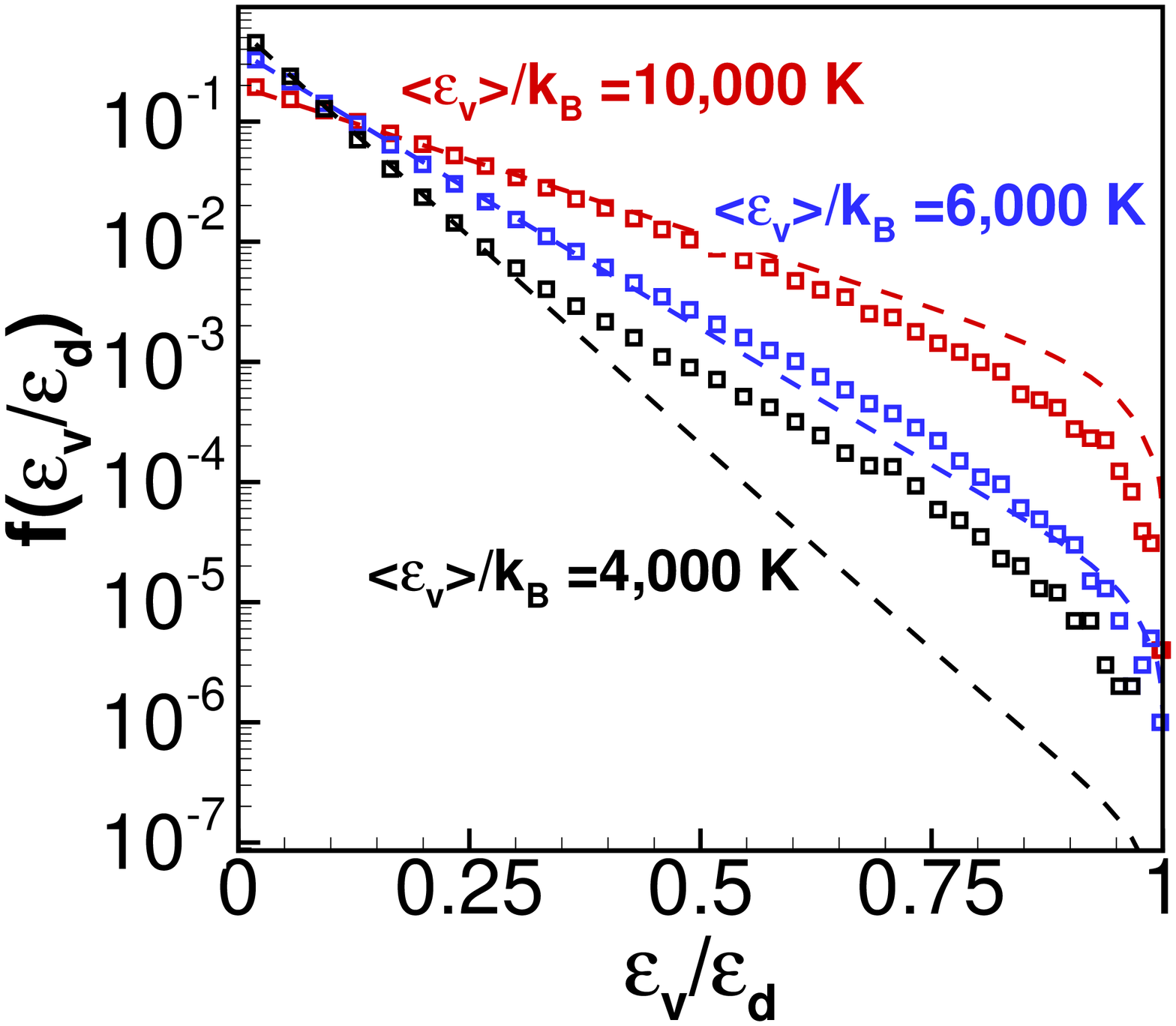}
\label{10K_distros_DMS}
}
 \caption{ (a)  Isothermal ro-vibrational excitation of oxygen gas from DMS \cite{grover2019direct} at $T = 10,000$ K.  (b) Corresponding vibrational energy distribution evolution during isothermal excitation. Symbols represent non-Boltzmann distributions and dashed lines represent Boltzmann distribution corresponding to the average vibrational energy. }
  \label{10K_DMS_Relax_QSS}
\end{figure}
\section{Non-Boltzmann Distributions Observed in Ab Initio Calculations}

\subsection{General nonequilibrium trends}

Direct molecular simulation (DMS) has been recently used to simulate post-shock representative conditions, specifically using zero-dimensional isothermal \cite{valentini2016dynamics,valentini2015direct2,valentini2014rovibrational,valentini2015N4,valentini2015N3,valentini2015direct,grover2019direct,grover2019jtht} and adiabatic \cite{torres2019direct} relaxation approaches using \textit{ab initio} PESs\cite{paukku2013global,paukku2014erratum}. 
Figure \ref{10K_DMS_Relax_QSS} shows an example DMS calculation of oxygen rovibrational excitation and dissociation \cite{grover2019direct}, where twelve \textit{ab initio} PESs act as the sole model input (twelve PESs are essential to describe O$_2$-O$_2$ and O-O$_2$ interactions in the ground electronic state including all spin and spatial degeneracies \cite{paukku2017potential,varga2017potentialO3,paukku2018potential}). The zero-dimensional system (in Fig.~\ref{10K_DMS_Relax_QSS}) is initialized with approximately 1 million O$_2$ molecules and the initial rovibrational populations are sampled from a Boltzmann distribution corresponding to $T=1000K$ (using the WKB method applied to the diatomic PES as described in Ref. \cite{bender2015improved}). To maintain isothermal conditions, the center-of-mass translational energies of the molecules are re-sampled from a Maxwell-Boltzmann distribution corresponding to T = 10,000 K at each timestep. These conditions mimic immediate post-shock conditions, where the internal energy of the gas excites towards a high relative translational energy and the gas begins to dissociate. 

The first notable trend, apparent from the average internal energies plotted in Fig.~\ref{10K_excitation_DMS}, is that significant dissociation occurs only after the vibrational energy of the gas has excited sufficiently; despite the rotational energy being almost fully excited. Second, the gas reaches a quasi-steady-state (QSS) where the internal energy (rotational and vibrational) is increasing towards the translational energy due to non-dissociative collisional processes, however, substantial dissociation reactions are simultaneously removing internal energy from the gas. These processes balance, leading to a plateau in the average internal energies. Although difficult to see in Fig.~\ref{10K_excitation_DMS} due to the logarithmic scale, the QSS region has been observed in numerous studies  \cite{valentini2016dynamics,valentini2015direct2,valentini2014rovibrational,valentini2015N4,valentini2015N3,valentini2015direct,grover2019direct,grover2019jtht, panesi2014pre,macdonald2018construction_DMS,macdonald2018_QCT}.

Figure \ref{10K_distros_DMS} shows the vibrational energy distribution functions at various times during the simulation; the focus of the current article. The main physical trends observed in the vibrational energy distribution functions are (a) the distributions are non-Boltzmann, (b) during initial excitation the high-energy tail of the distribution is over-populated compared to the corresponding Boltzmann distribution and, (c) during QSS the high-energy tail is depleted relative to the corresponding Boltzmann distribution. Such depleted vibrational energy distributions have been shown to reduce computed dissociation rate constants by a factor of 3-5 for nitrogen \cite{bender2015improved,valentini2016dynamics}, and by a factor of 2-5 for oxygen \cite{grover2019direct}, compared to computed dissociation rates that assume equilibrium (Boltzmann) internal energy distributions \cite{bender2015improved,chaudhry2018qct}. 
\subsection{Implications for Modeling}
As evident from Fig. \ref{10K_distros_DMS}, the vibrational energy distributions exhibit overpopulation of high energy states (compared to Boltzmann) during the initial transient period associated with rapid vibrational energy excitation. As the gas relaxes and begins to dissociate, the high energy states transition from being overpopulated to being depleted. Although at some specific average vibrational energy the distribution appears close to a Boltzmann distribution, the vibrational energy distributions are clearly non-Boltzmann throughout the entire simulation. 

Focusing on the distribution functions during the transient phase (rapid excitation phase where $\langle \epsilon_v \rangle / k_B \leq 6000 K$), the populations of low vibrational energy states are accurately described by a Boltzmann distribution (say $f_0(v;T_v)$) corresponding to the average vibrational energy (i.e $ \sum_v \epsilon_v(v) f_0(v;T_v)  = \langle \epsilon_v \rangle$). This implies that a model for the distribution function should clearly have dependence on the local average vibrational energy of the gas, equivalently on $T_v$ which represents the temperature that recovers the average vibrational energy using a Boltzmann distribution. It is straightforward that the mapping between $T_v$ and $\langle \epsilon_v \rangle$ is bijective and, therefore, the distribution for a given $\langle \epsilon_v \rangle$ is unique.
 \begin{figure}
   \subfigure[]{
   \includegraphics[width=3.0in,trim={0.05cm 0.05cm 0.0cm 1.0cm},clip]{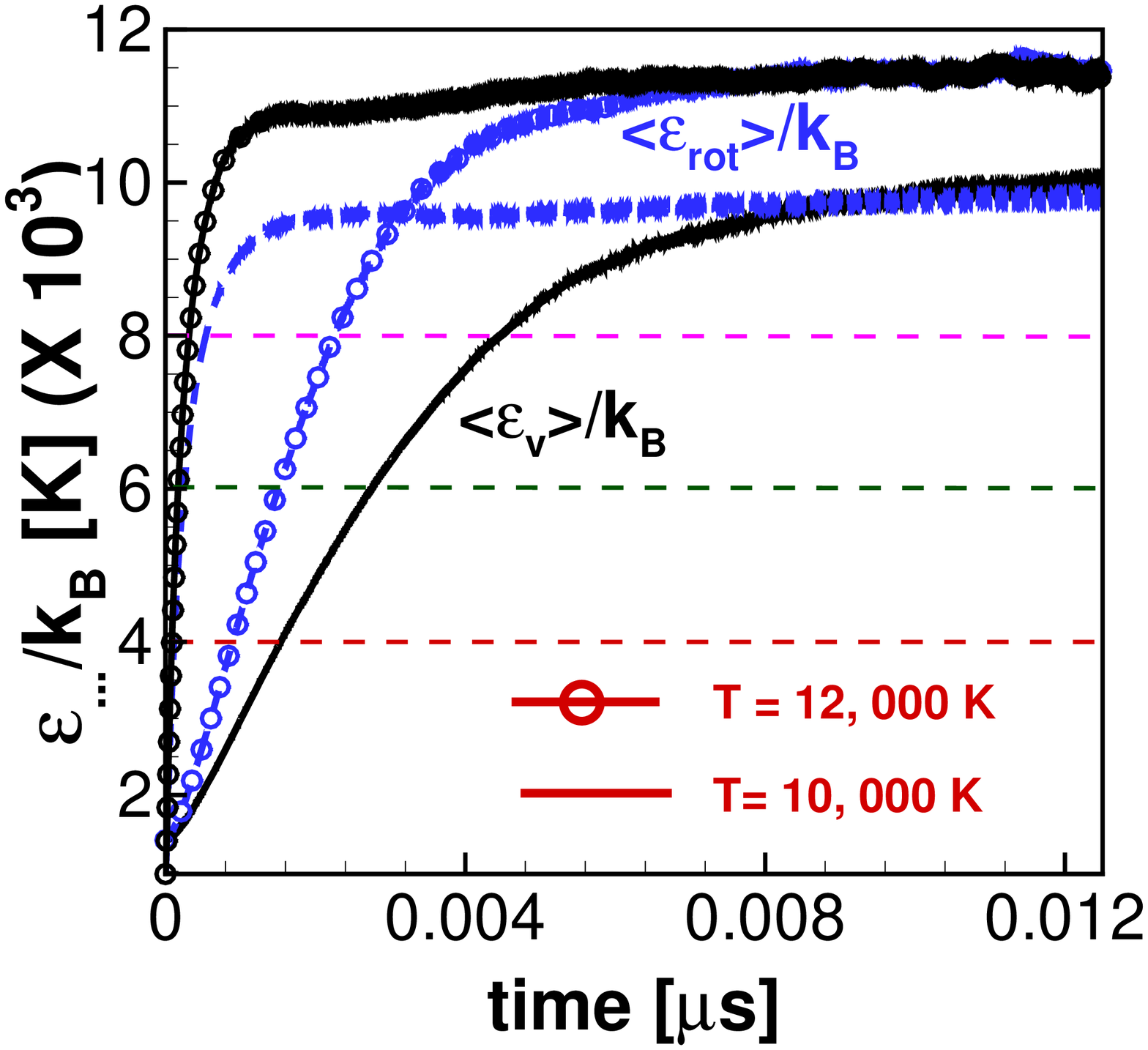}
    \label{excitation_10K_12K_DMS}
  }
\subfigure[]{
\includegraphics[width=3.0in,trim={0.25cm 0.25cm 0.0cm 2.0cm},clip]{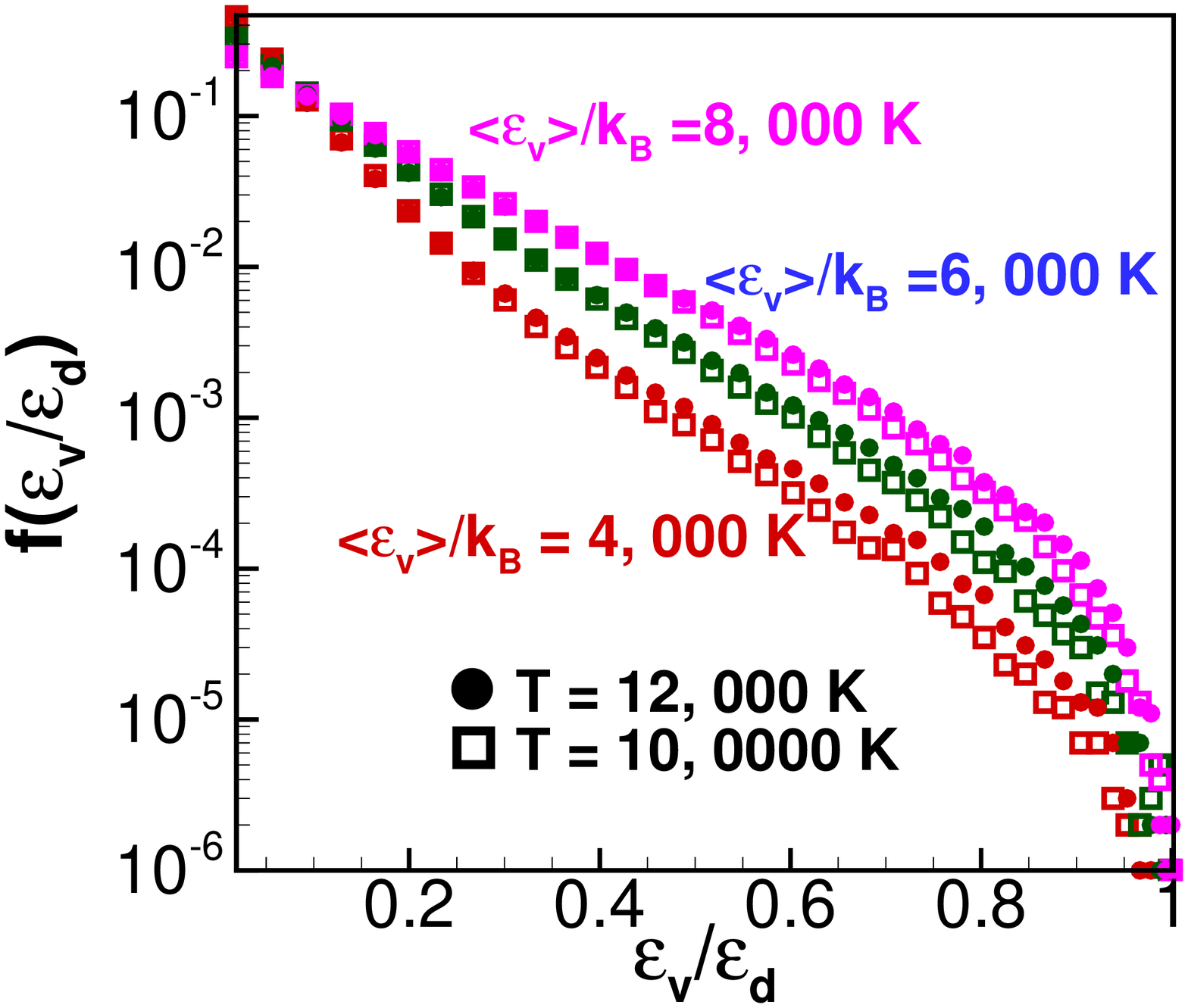}
 \label{distros_10K_12K_DMS}
}
 \caption{ 
(a)  Isothermal ro-vibrational excitation of oxygen gas from DMS \cite{grover2019direct} at $T = 10,000$ K (shown with circular symbols) and $12, 000K$ (shown in Black)  (b) Corresponding vibrational energy distribution evolution during isothermal excitation .  }
  \label{10K_12K_DMS}
\end{figure}
However, the populations in higher states deviate significantly from Boltzmann. Since the probability of dissociation has an exponential dependence on the vibrational energy of the dissociating molecule, such deviations in the high energy populations can have a significant effect on the overall dissociation rate in the gas. Therefore, an additional dependence is required for the model to accurately capture the populations of high energy states. Furthermore, we find that the extent of overpopulation is dependent on the translational temperature. 
 This is shown in Fig. \ref{10K_12K_DMS}, where an isothermal relaxation simulation with a fixed translational temperature of 12,000 K is compared to the 10,000 K simulation from Fig. \ref{10K_DMS_Relax_QSS}. Figure \ref{excitation_10K_12K_DMS} shows that the trends in average rotational and vibrational energies are similar for the two conditions; initial rapid excitation followed by a QSS region. Figure \ref{distros_10K_12K_DMS} plots the vibrational energy distributions from the 12,000 K case at instants where the average vibrational energy corresponds to $\langle \epsilon_v \rangle / k_B = 4000 K$ and $\langle \epsilon_v \rangle / k_B = 6000 K$, and directly compares them to the results from the 10,000 K case. It is observed that during the transient phase the distribution functions for different translational temperatures are not the same. This implies that the distribution functions in the transient phase are dependent on both the local average vibrational energy and also the temperature towards which the system is exciting. A unique representation of the nonequilibrium \textit{state} \footnote{strictly speaking state should only be used for thermodynamic equilibrium, here we use the term loosely} in the transient phase, therefore, requires the dependence on the temperature $T$, on the local average vibrational energy and on the initial condition. 
 
 Focusing on the depleted vibrational energy distribution functions in QSS, prior research \cite{grover2019jtht} has shown that the QSS distributions are independent of the internal energy initial conditions, rather the extent of depletion only depends on the translational temperature (in an isothermal heat bath). 
This dependence was recently studied by Singh and Schwartzentruber \cite{Singhpnas} using a framework called surprisal analysis \cite{levine2009molecular,levine1978information,levine1971collision}, and elements of this work will be used in the general model described in the next section.

Before describing the new generalized model for nonequilibrium vibrational energy distributions, it is relevant to discuss prior modeling efforts.
The evolution of vibrational energy distributions under isothermal conditions has been studied, analytically, for harmonic oscillators in the absence of dissociation. Specifically, the study used Landau-Teller transition probabilities \cite{rubin1956relaxation} assuming only mono-quantum transitions. It is noted that, for Landau-Teller transition probabilities with mono-quantum transitions and the harmonic oscillator assumption, the gas evolves as series of Boltzmann distributions corresponding to $T_v$. However, if exponential transition probabilities \cite{rubin1956relaxationprob} are assumed, the distributions also become a function of the initial state. Such analysis was extended to anharmonic oscillators \cite{bazley1958studies}, and dissociation effects were incorporated by allowing dissociation from only the last quantum state \cite{montroll1957application}. While these studies serve as useful background for vibrational energy relaxation physics, the assumptions that transitions occur only between nearby states and that dissociation occurs only from the last quantum state are not valid at the high temperatures of interest to our study. Most importantly, such analysis cannot be extended to post-shock conditions where the translational temperature is not constant. In the next section we describe a generalized model that accurately captures non-Boltzmann effects through dependence on T, $T_v$, and initial condition.

 \section{Model for Non-Boltzmann Vibrational Energy Distributions}
 



In this section, we present an analytical model for non-Boltzmann vibrational energy distributions. The general form includes a contribution due to overpopulation and a contribution due to depletion with an overall dependence on both $T$ and $T_v$, as well as the initial condition.  

In prior work, a surprisal framework by Singh and Schwartzentruber \cite{Singhpnas} was used to model the depletion of high energy states due to QSS dissociation. If $f_0$ represents a Boltzmann distribution corresponding to the average vibrational energy, then the non-Boltzmann QSS distribution is modeled as: 
  \begin{equation}
\begin{split}
f_{\QSS} (v) = \cfrac{f_0 \exp\left[-\lambda_{1,v} \frac{\langle \epsilon_t \rangle}{\epsilon_d} v \right]} { \sum\limits_{k}^{v_{\mx}}  f_0 \exp\left[-\lambda_{1,v} \frac{\langle \epsilon_t \rangle}{\epsilon_d} k \right]} 
\end{split}
\label{fdepleted}
\end{equation}
where $\lambda_{1,v}$ is a free parameter chosen to match the depletion observed in DMS calculations \cite{Singhpnas}.

In-fact, the complete model proposed by the authors in earlier work is generalized for rotational and vibrational energy distributions, and also includes a term controlling overpopulation of energy states:
\begin{equation}
    -\log\left[\frac{f(i)}{f_0 (i;T_i)}\right]=\lambda_{0}+\lambda_{1,i} \frac{\langle \epsilon_t \rangle}{\epsilon_d} i+\lambda_{2,i} \left( \frac{2}{3}\frac{\langle \epsilon_t \rangle}{\langle \epsilon_i \rangle} - \frac{3}{2}\frac{\langle \epsilon_i \rangle}{\langle \epsilon_t \rangle} \right)^\psi i~.
    \label{ned_model_v}
\end{equation}
Here $\log\left[f (i)/f_0 (i; T_i) \right]$ is the surprisal, $i$ can be rotational ($j$) or vibrational ($v$) quantum state, $f_0 (i;T_i)$ is a Boltzmann distribution corresponding to $T_i$, $\lambda_0$ ensures $\sum_i f(i)$ equals unity, $\lambda_{2,i}$ and $\psi$ (odd integer) are free parameter constants chosen to match overpopulation trends observed in DMS calculations, and the term involving $\lambda_{1,i}$ is the QSS depletion term described above. 
The overpopulation term dominates in the transient phase involving rapid excitation ($T>>T_v$), whereas the depletion term dominates in the QSS phase ($T\approx T_v$). 

In this article we develop a model that includes three substantial improvements not considered in the previous work. Specifically, (i) dependence on the initial state is now included, (ii) the new model ensures that the moment of the distribution recovers the correct average vibrational energy ($\langle \epsilon_v \rangle$), and (iii) the new model no longer requires any free parameters for the terms controlling overpopulation ($\lambda_{2,i}$ and $\psi$ are no longer required). 

The proposed model for non-Boltzmann vibrational energy distributions, $f^{NB}(v)$, behind strong shocks is given by:
\begin{equation}
     f^{NB}(v) = \cfrac{\tilde{f}(v;T_v,T_0)+\Lambda f^d(v;T)}{1+\Lambda} 
     \label{fnb_overall}
\end{equation}
 where $T_0$  is the reference (initial state) temperature and $\tilde{f}(v;T_v,T_0)$ and $f^d(v;T)$ are defined as follows:
 \begin{equation}
 \begin{split}
     f^d(v;T) = \cfrac{\exp\left[-\cfrac{\epsilon_v(v)}{k_B T}-\lambda_{1,v} \cfrac{\langle \epsilon_t \rangle}{\epsilon_d} v\right]}{ \sum_v \exp\left[-\cfrac{\epsilon_v(v)}{k_B T}-\lambda_{1,v} \cfrac{\langle \epsilon_t \rangle}{\epsilon_d} v\right]}
     \label{depleted_distros}
     \end{split}
 \end{equation}
 
  \begin{equation}
 \begin{split}
       \tilde{f}(v;T_v,T_0) = \cfrac{\exp\left[-\cfrac{\Delta_{\epsilon} v}{k_B T_v}-\cfrac{\Delta_{\epsilon} v -\epsilon_v(v)}{k_B T_0}-\lambda_{1,v} \cfrac{\langle \epsilon_t \rangle}{\epsilon_d} v\right]}{ \sum_v \exp\left[-\cfrac{\Delta_{\epsilon} v}{k_B T_v}-\cfrac{\Delta_{\epsilon} v -\epsilon_v(v)}{k_B T_0}-\lambda_{1,v} \cfrac{\langle \epsilon_t \rangle}{\epsilon_d} v\right]}
       \label{overpop_distros}
        \end{split}
 \end{equation}
 where $\Delta_{\epsilon} = \epsilon_v(1)-\epsilon_v(0)$. In Eq.~\ref{overpop_distros}, the low energy states which have nearly constant energy spacing ($\Delta_{\epsilon}$) are populated as Boltzmann distribution such as in the simple harmonic oscillator assumption and the high energy states are depleted due to dissociation. The mid term in Eq.~\ref{overpop_distros} adds dependence on the initial condition ($T_0$), which has vanishing contribution in the low lying states. 
 
 In Eq.~\ref{overpop_distros}, the parameter $\Lambda$ ensures that $ \sum_v \epsilon_v(v) f^{NB}(v)  = \langle \epsilon_v \rangle$ i.e
\begin{equation}
\begin{split}
   \sum_v  \epsilon_v(v) f^{NB}(v) = \cfrac{\sum_v  \epsilon_v(v)\tilde{f}(v;T_v,T_0)+\Lambda \sum_v  \epsilon_v(v) f^d(v;T)}{1+\Lambda} 
   \\
   \langle \epsilon_v \rangle = \cfrac{\tilde{\langle \epsilon_v \rangle}+\Lambda \langle \epsilon_v \rangle^*}{1+\Lambda} \hspace{0.0in}
   \implies \Lambda = \cfrac{\tilde{\langle \epsilon_v \rangle}-\langle \epsilon_v \rangle}{\langle \epsilon_v \rangle-\langle \epsilon_v \rangle^*}
   \end{split} 
     \label{enb_overall}
\end{equation}
where $\langle \epsilon_v \rangle^*$ is average vibrational energy in QSS state and $\tilde{\langle \epsilon_v \rangle}$ is the dominant contribution of the low lying energy states to the average vibrational energy. 

The model therefore has two limits. An approximate Boltzmann distribution corresponding to the reference temperature $T_0$ (initial condition) is recovered in the limit of $\Lambda = 0$ ($T_v \approx T_0$) , and in the limit of $\Lambda^{-1}=0$ ($T_v \approx T$) the model distribution reduces to a depleted QSS distribution (equivalent to that used in previous work: $f_{\QSS} (v)$ in Eq. \ref{fdepleted}). Between these limits, in the transient phase corresponding to rapid vibrational energy excitation, the term $\tilde{f}(v;T_v,T_0)$ (Eq.~\ref{overpop_distros}) contributes and contains three main terms: a Bolztmann distribution based on $T_0$ , a Boltzmann based on $T_v$ and a  depleted distribution for the high energy states. The  overpopulation of the high energy tails is captured by Eq.~\ref{overpop_distros}, extent of which depends on the parameter $\Lambda$. 

Finally, we have found that under many conditions, the reference temperature term is only necessary for the region immediately behind the shock-front where the vibrational energy distribution still has populations from the initial state. This shock-front region is typically not resolved in continuum simulations and therefore, for practical purposes, the reference state (parameter $T_0$) may be omitted depending how the model will be applied. Under this assumption, the non-Boltzmann vibrational distribution model reduces to:
  \begin{equation}
 \begin{split}
       \tilde{f}(v;T_v,T_0) \approx \cfrac{\exp\left[-\cfrac{\Delta_{\epsilon} v}{k_B T_v}-\lambda_{1,v} \cfrac{\langle \epsilon_t \rangle}{\epsilon_d} v\right]}{ \sum\limits_{v}^{v_{\mx}}  \exp\left[-\cfrac{\Delta_{\epsilon} v}{k_B T_v}-\lambda_{1,v} \cfrac{\langle \epsilon_t \rangle}{\epsilon_d} v\right]}
        \end{split}
 \end{equation}
 
In summary, an important aspect of the nonequilibrium vibrational energy distribution model in Eq. \ref{fnb_overall} is that it is formulated in terms of average molecular energies, and can therefore be evaluated within a multi-temperature CFD simulation (a function of $T$ and $T_v$). The distribution is bimodal and combines populations based on the (local) average vibrational energy and the local average translational energy (to which the distribution is relaxing towards). The precise combination (and therefore degree of overpopulation) is controlled by $\Lambda$, which is analytically known by enforcing that the moment of the distribution recovers the correct average vibrational energy. Finally, the depletion of high vibrational levels must be accurately modeled in each term of the formulation, using the surprisal formulation used in prior work \cite{Singhpnas}, since such high energy levels are strongly coupled to dissociation. In the next section, this simple model will be tested against DMS simulation results.

\section{Model Verification Against DMS Results}
In this section, the new model is used to estimate the nonequilibrium vibrational energy distribution functions for both isothermal conditions and adiabatic conditions for both oxygen and nitrogen gases, and the results are compared to baseline DMS calculations to assess the accuracy of the model.

\subsection{Isothermal relaxation in nitrogen and oxygen}
 \begin{figure}
  \subfigure[]
  {
   \includegraphics[width=3.4in,trim={0.15cm 0.15cm 0.15cm 1.75cm},clip]{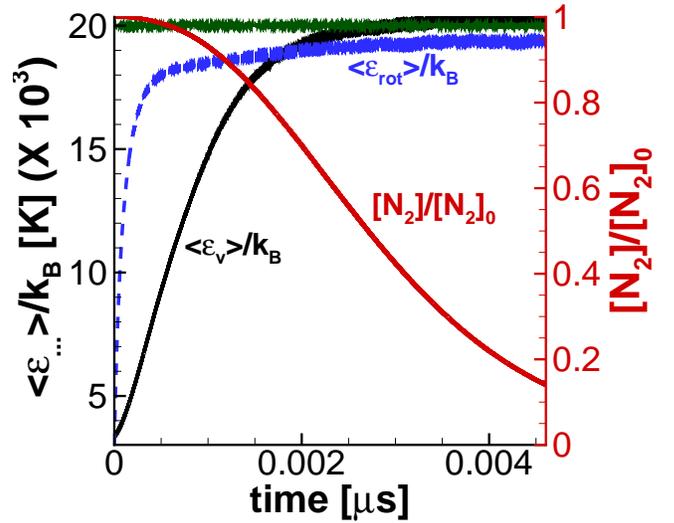}
    \label{Nitrogen_Comp_Isothermal}
    }
    \subfigure[]{
   \includegraphics[width=3.0in,trim={0.15cm 0.65cm 0.15cm 0.15cm},clip]{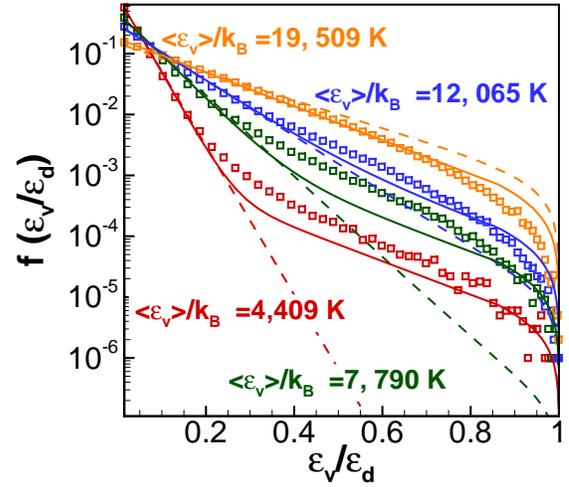}
    \label{Nitrogen_Distros_Isothermal}
  }
 \caption{ 
 (a)  Isothermal ro-vibrational excitation of oxygen gas from DMS \cite{valentini2014rovibrational} at $T = 20,000$ K (b)Vibrational energy distributions from the model (solid lines), DMS (symbols), and Boltzmann distributions (dashed line).}
  \label{Isothermal_Nitrogen_20K_DMS_Model}
\end{figure}

The first test case involves an isothermal relaxation calculation in nitrogen gas. One million nitrogen molecules are initialized with low average rotational $(\langle \epsilon_{rot} \rangle =3, 000 \ k_B \ K$ and vibrational $(\langle \epsilon_{v} \rangle =3, 370 \ k_B \ K)$ energy, while the translational energy is maintained at $T=20, 000$ K by resampling center-of-mass velocities from a Maxwell-Boltzmann distribution after each simulation timestep. These DMS results are taken from Ref. \cite{valentini2016dynamics}, which includes full details of the simulation procedure and full analysis of the results. The DMS calculation uses the nitrogen PES described in Ref. \cite{bender2015improved,paukku2013global} for both N$_2$-N$_2$ and N-N$_2$ collisions as the sole model input. Evolution of the average rotational and vibrational energies, as well as the mole fraction of N$_2$ is shown in Fig.~\ref{Nitrogen_Comp_Isothermal}. As the gas evolves, the vibrational energy distributions from DMS  are plotted, along with the corresponding Boltzmann vibrational energy distribution functions, at several instances each denoted by the instantaneous average vibrational energy of the gas, in Fig.~\ref{Nitrogen_Distros_Isothermal}. During the rapid excitation phase, the Boltzmann distributions significantly under-predict the population of high vibrational energy states when compared with the DMS distributions. Therefore, the actual dissociation rate during this phase may be significantly higher compared to a dissociation rate evaluated by averaging over a Boltzmann distribution corresponding to the local $T_v$. During the QSS phase, the Boltzmann distribution over-predicts the population of high vibrational energy states, and would over-predict the true dissociation rate. Results using the new model are also shown in Fig.~\ref{Nitrogen_Distros_Isothermal}. Although there are differences between the model and DMS results, considering the simplicity of the model, it succeeds in accurately capturing the non-Boltzmann overpopulation and depletion observed in the DMS calculation. 
 \begin{figure}
   \includegraphics[width=3.2in,trim={0.15cm 0.15cm 0.15cm 1.75cm},clip]{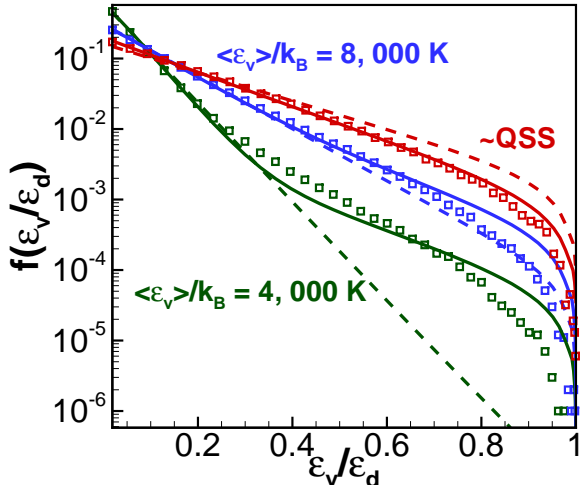}
 \caption{
Vibrational energy distribution evolution for isothermal ro-vibrational excitation of oxygen gas from DMS \cite{grover2019direct} at $T = 12,000$ K. Symbols represent non-Boltzmann distributions and dashed lines represent Boltzmann distribution corresponding to the average vibrational energy.}
  \label{12K_excitation_DMS_distros}
\end{figure}
The next two test cases involve isothermal relaxation calculations in oxygen gas at $T=12,000$ K and $T=10,000$ K. The DMS calculations used three PESs for O$_2$-O$_2$ \cite{paukku2018potential} collisions and nine PESs for O-O$_2$ collisions \cite{varga2017potentialO3}, which account for all spin-couplings and spatial degeneracies for collision involving oxygen in the ground electronic state. Full details of the DMS calculations and full analysis of results can be found in Ref. \cite{grover2019direct,grover2019jtht}. 

The trends in average vibrational energy and the composition for the $T=12, 000$ K case were shown previously in Fig.~\ref{10K_12K_DMS}. The vibrational distribution functions during the rapid excitation phase, for the $T=12,000$ K case and during the approach to QSS are shown in Figure~\ref{12K_excitation_DMS_distros}. Similar to the nitrogen results, the Boltzmann distributions corresponding to the instantaneous average vibrational energy significantly under-predict the populations of high vibrational energy levels initially, however, over-predict the populations as the gas reaches QSS. The accuracy of the new model for this oxygen case is similar to the accuracy found for nitrogen, where both the overpopulation and depletion of high energy states is captured.

\begin{figure}
   \includegraphics[width=3.2in,trim={0.05cm 0.05cm 0.0cm 0.05cm},clip]{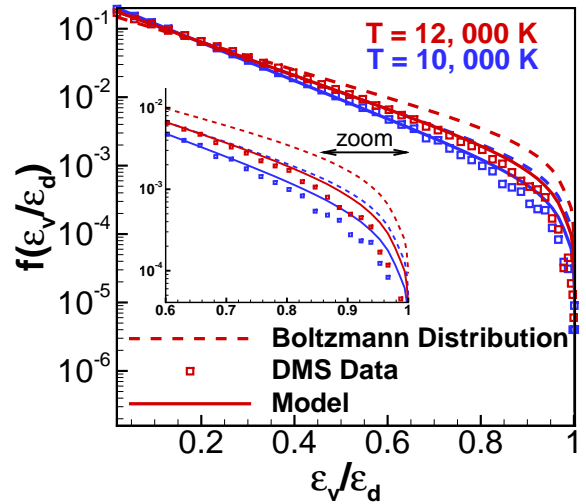}
    \label{QSS_10K_DMS_Model}
 \caption{ 
 Vibrational energy distributions in QSS from the model, DMS\cite{grover2019direct}, and Boltzmann distributions at $T = 10,000$ K  and $12, 000$ K .}
  \label{QSS_10K_12K_DMS_Model}
\end{figure}

Figure~\ref{QSS_10K_12K_DMS_Model} shows the prediction of non-Boltzmann distributions only in the QSS phase for the isothermal oxygen relaxations at $T=12,000$ K and also at $T=10, 000$ K. The depleted distributions predicted by the model are seen to be in good agreement with the QSS distributions obtained in the DMS calculations.

At this point, it is important to note that the only free parameter (i.e. fitting parameter) necessary for the new nonequilibrium vibrational energy distribution model is $\lambda_{1,v}$. This is a constant parameter, and in prior work for nitrogen, a value of $\lambda_{1,v} = 0.08$ was found to accurately reproduce the degree of depletion in high energy states observed in DMS calculations. We have used the same value of $\lambda_{1,v}$ in all of the results (both for nitrogen and oxygen) presented in this article. The reason that a simple constant is applicable to different gases is because the dependence $\langle \epsilon_{t} \rangle / \epsilon_d$ is general to different molecules with different dissociation energy barriers, enabling a constant coefficient to control the overall ``amount'' of depletion, which we find is rather consistent between oxygen and nitrogen. The only other parameter that may be required is the reference state temperature. For the results presented in this article we use $T_0 = 300$ K. Based on experience, we find the distribution results are not sensitive to the precise value of $T_0$, and therefore we recommend a value of 300 K for vibrationally cold free-stream conditions (flight conditions or some expansion tunnel conditions), however, for reflected shock-tunnel conditions where the free-stream vibrational temperature may be frozen at a high value, one should set $T_0$ to the free-stream (pre-shock) vibrational temperature.

\subsection{Adiabatic relaxation in nitrogen and oxygen}
In this subsection, we compare the new model predictions with more recent DMS calculations under adiabatic conditions (constant total energy). Adiabatic conditions are more representative of post-shock conditions where the translational temperature of the gas drops quickly as energy is transferred into rotational and vibrational energy and where translational and internal energy is continually removed due to dissociation reactions. DMS calculations for both oxygen and nitrogen gases are presented.
 \begin{figure}
     \subfigure[]{
  \includegraphics[width=3.4in,trim={0.15cm 0.15cm 0.05cm 2.25cm},clip]{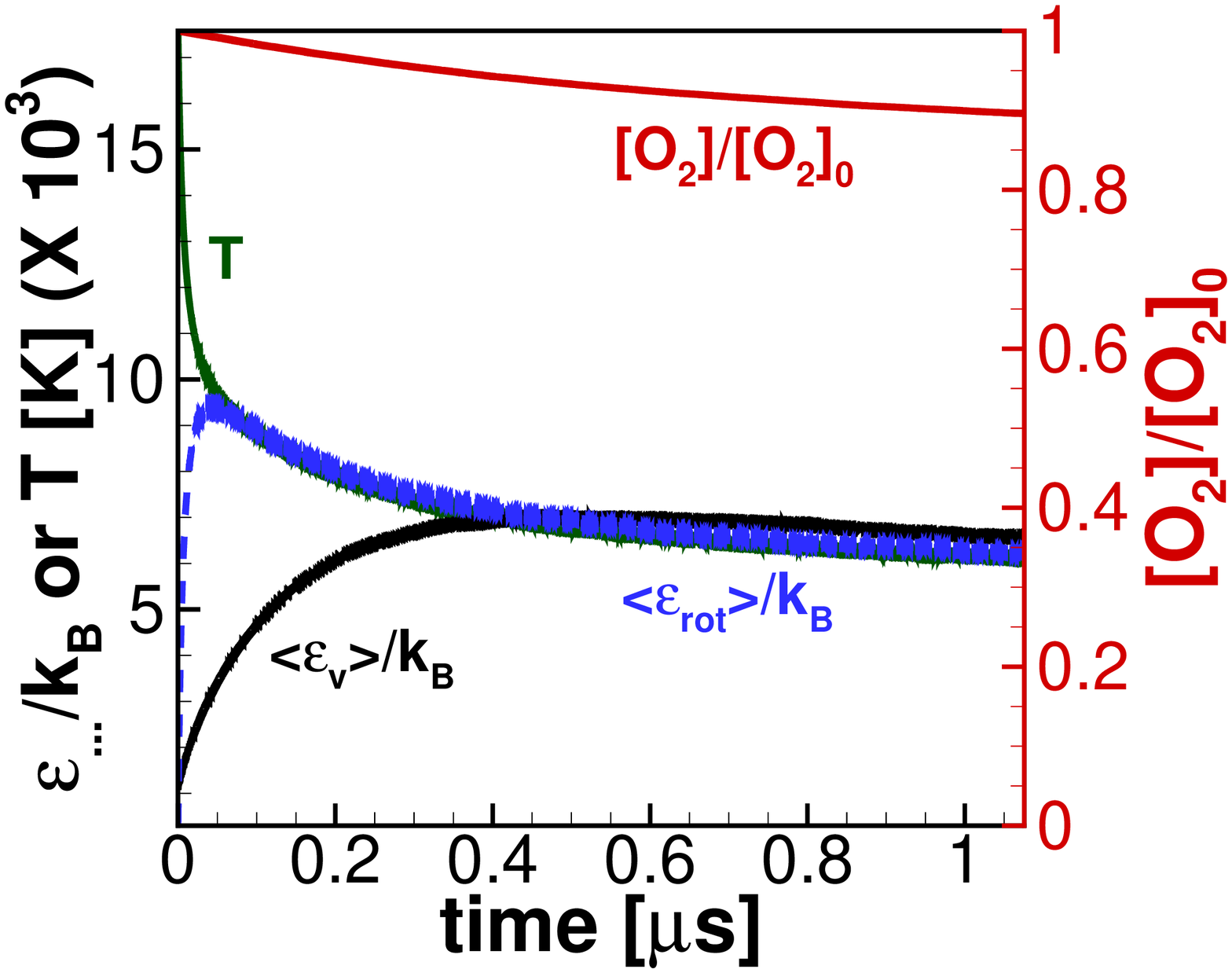}
    \label{Oxygen_comp_adiabatic}}
       \subfigure[]{
          \includegraphics[width=3.0in,trim={0.15cm 0.15cm 0.15cm 1.40cm},clip]{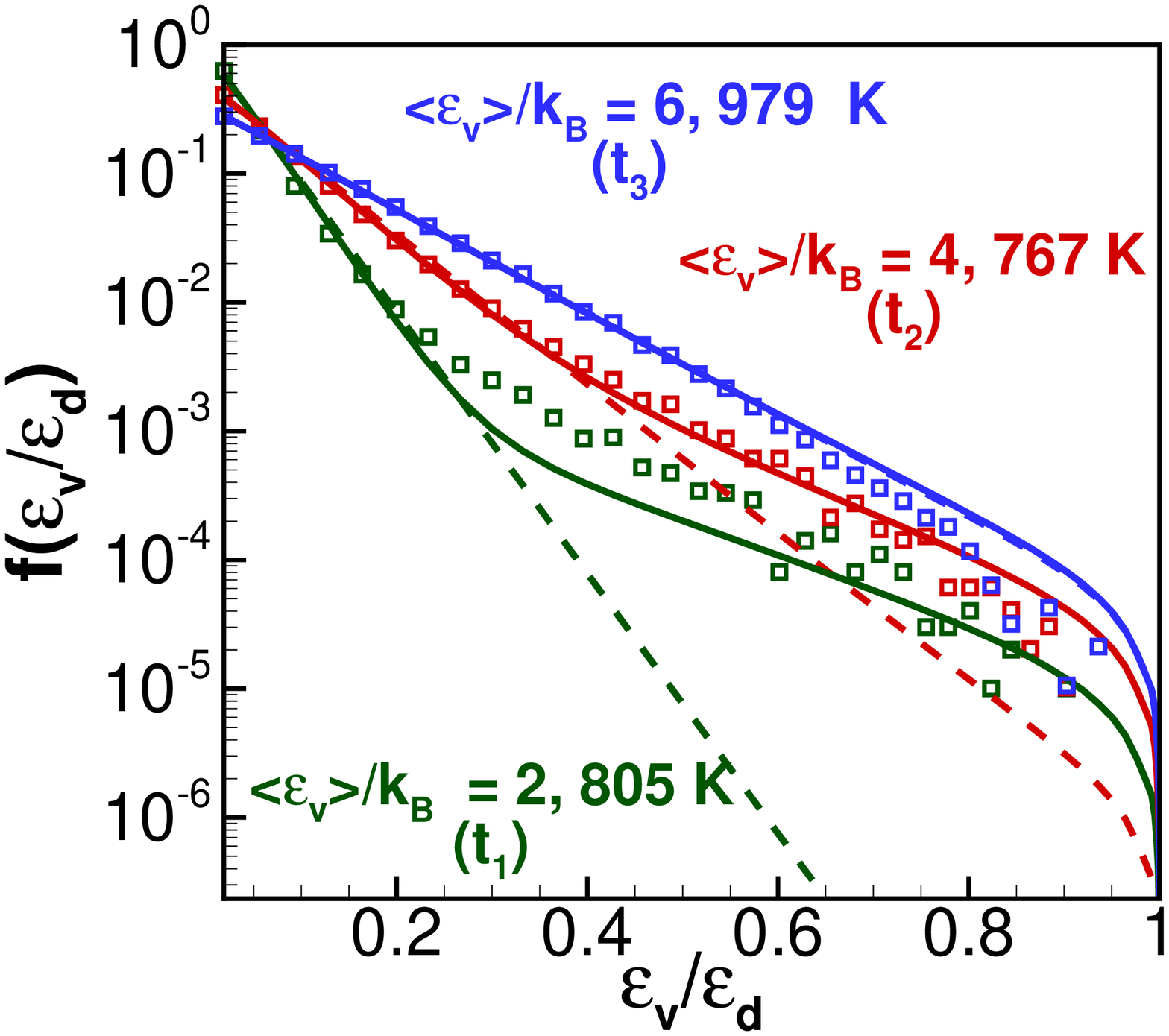}
    \label{Oxygen_distros_adiabatic}
  }
  \caption{ 
 (a) Adiabatic ro-vibrational excitation of oxygen gas from DMS \cite{torresoxygen2019}  (b) Evolution of vibrational energy distributions using DMS (symbols) and model (solid lines) along with Boltzmann distributions (dashed lines) where $t_1= 0.028 \mu$s, $t_1= 0.100 \mu$s $t_1= 0.3960 \mu$s}
  \label{Oxygen_adiabatic}
\end{figure}
For the oxygen case, one hundred thousand O$_2$ molecules, corresponding to a density of $0.0230$ kg/m$^3$, are initialized at a high translational energy T $ = 17, 578 K$ and internal energies are sampled from a Boltzmann distribution at $300$ K. In contrast to the isothermal DMS calculations, the post-collision center-of-mass velocities are maintained (they are not resampled from Maxwell-Boltzmann distributions) and are used as the pre-collision properties for the molecule's (or atom's) next collision. In this manner, total energy is conserved and the gas relaxes in an adiabatic manner. Evolution of the average rotational and vibrational energy, as well as the molar fraction of molecular oxygen, is shown in Fig.~\ref{Oxygen_adiabatic} along with the translational temperature. Under these conditions, the rotational energy rapidly equilibrates with the translational energy, compared to the vibrational energy which excites at a much slower rate. The vibrational energy distributions at different time instants are shown in Fig.~\ref{Oxygen_distros_adiabatic}, labelled by the instantaneous average vibrational energy of the gas. The DMS data is taken from Ref.\cite{torresoxygen2019}.

\begin{figure}
    \subfigure[]{
  \includegraphics[width=3.5in,trim={0.40cm 0.75cm 0.05cm 1.25cm},clip]{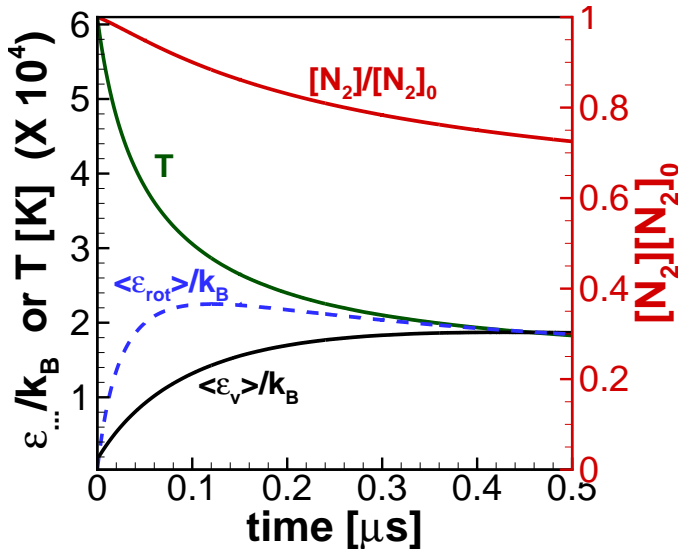}
    \label{N4_comp_adiabatic}}
    \subfigure[]{
   \includegraphics[width=2.9in,trim={0.15cm 0.35cm 0.5cm 0.45cm},clip]{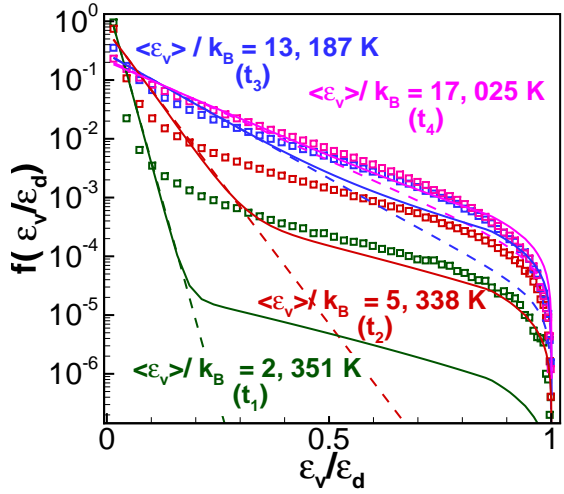}
    \label{N4_distros_adiabatic}
  }
 \caption{ 
 (a) Adiabatic ro-vibrational excitation of nitrogen gas from DMS \cite{torresoxygen2019}  (b) Evolution of vibrational energy distributions using DMS (symbols) and model (solid lines) along with Boltzmann distributions (dashed lines), where $t_1 =0.003 \mu s, t_2 =0.02 \mu s, t_3 = 0.1 \mu s$ and $t_4 = 0.2 \mu s$}
  \label{N4_adiabatic}
\end{figure}

For the nitrogen case, approximately five million nitrogen molecules, corresponding to a density of $0.0298$ kg/m$^3$, are initialized with high translational energy ($T=60, 000$ K), whereas, rotational and vibrational energies are sampled from a Boltzmann distribution corresponding to $300 $ K. Evolution of the average energies and molar fraction of N$_2$ is shown in Fig.~\ref{N4_comp_adiabatic}.  The vibrational energy distributions at various time instants are shown in Fig.~\ref{N4_distros_adiabatic}. The data is taken from Ref.\cite{torres2019direct}.

The trends in average energies and mole fractions of diatomic species are similar for both oxygen and nitrogen cases. The vibrational energy distribution functions for both oxygen and nitrogen cases show similar trends during the gas evolution, and these trends are also similar to those seen in the isothermal simulations. Specifically, compared to Boltzmann distributions, the vibrational energy distributions exhibit overpopulation of high energy states during the rapid excitation phase and depletion during the QSS dissociation phase. The nitrogen case has considerably more severe initial conditions ($T(t=0) = 60,000$ K), leading very substantial overpopulation in the earliest stages of excitation (as seen in Fig.~\ref{N4_distros_adiabatic}). In general, the degree of overpopulation is more significant under adiabatic conditions compared to isothermal conditions (for the same QSS temperature), since the translational temperatures are initially very high in the adiabatic case. The degree of depletion during QSS is similar between isothermal and adiabatic simulations (again for the same QSS temperature). In-fact, this result has been confirmed in a recent study where internal energy distribution functions were directly compared between isothermal and adiabatic conditions using DMS \cite{torres2019direct}. It was found that under adiabatic conditions, the gas relaxes as a series of QSS (depleted) distributions that depend on the instantaneous translational temperature (which slowly decreases due to continued dissociation as seen in Figs. \ref{Oxygen_comp_adiabatic} and \ref{N4_comp_adiabatic}. For the same translational temperature, the depleted distributions found under adiabatic relaxation conditions were virtually identical to those found under isothermal relaxation conditions. This is one reason why the new model presented in this article is accurate for both isothermal and adiabatic relaxation conditions. 

As shown in Figs. \ref{Oxygen_distros_adiabatic} and \ref{N4_distros_adiabatic}, the model accurately captures both the overpopulation and depletion of high vibrational energy states at various instances during adiabatic relaxations for both oxygen and nitrogen. The apparent disagreement in the QSS regions for oxygen (the $\langle \epsilon_{v} \rangle = 6979 k_B$ curve in Fig.~\ref{Oxygen_distros_adiabatic}) is in-part due to statistical scatter in the DMS calculation results, where only a few molecules were sampled in the highest energy levels. The largest disagreement between DMS and the new model is found only in the earliest stages of rapid excitation for the nitrogen case where the translational temperature is upwards of 50,000 K for a very short time.

\section{Evaluating Continuum Dissociation Rate Constants}\label{Rates_from_state_specific_rates}
A continuum-level nonequilibrium dissociation rate constant can be calculated by the general expression,
\begin{equation}
\begin{split}
    k = \sum\limits_{v}^{v_{\mx}}   \sum\limits_{j}^{j_{\mx}(v)}  k(j, v; T) f^{NB}(j,v) \hspace{1.0 in},~
    \label{rate_eqn}
\end{split}
\end{equation}
where $k(j,v;T)$ are the state-specific dissociation rates corresponding to a translational temperature $T$ (assumed to be Maxwell-Boltzmann), which must be summed (integrated) over the local populations of rotational and vibrational quantum states ($j,v$) in the gas, written generally as non-Boltzmann distributions, $f^{NB}(j,v)$.
 \begin{figure}
 \centering
   \includegraphics[width=3.0in,trim={0.25cm 0.25cm 0.0cm 2.0cm},clip]{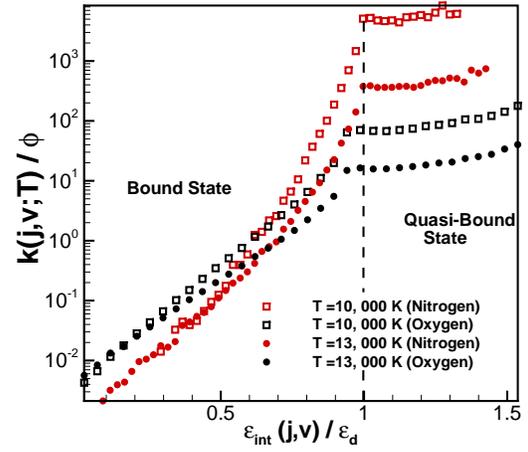}
    \label{Distros_20K_N4_DMS_model}
 \caption{ 
 Scaled state-specific dissociation rate constant ($k(j,v;T)/\phi$) variation with normalized $\epsilon_{int}(j,v)$. The factor $\phi = \sigma_{ref} \exp\left(-\epsilon_d/k_B T\right) \sqrt{8 k_B T/\pi \mu}$), where $\sigma_{ref}$ is reference cross-section and $\mu$ is reduced mass of reactants.  The data is obtained from QCT calculations for N$_2$+N$_2$\cite{bender2015improved} and  O$_2$+O$_2$\cite{ross2019}. }
    \label{N4_O4_probability}
\end{figure}

The state-specific dissociation rates, $k(j,v;T)$, can be determined for each quantum state using QCT analysis and compiled into databases \cite{andrienko2018vibrational,andrienko2017state,andrienko2016thermal,panesi2014pre,mankodi2018dissociation,mankodi2018cross, mankodi2018collision}. In many cases, well-defined trends have been observed in these state-specific rates and cross-sections, that can be fit accurately with simple functional forms \cite{singhmodeldevelopment2019}. For example, Fig. \ref{N4_O4_probability} plots the variation of the scaled dissociation rates ($k(j,v;T)/\phi$) versus the internal energy of the molecule for both oxygen and nitrogen at two different temperatures ($T=10, 000$K and $T=13, 000$K). The scaling factor corresponds to the Arrhenius term $\exp (\epsilon_d/ k_B T)$. These clear trends can be modeled using simple analytical expressions, for example using exponential functions as proposed in \cite{singh2017coupled}, or the proposed in Bias model \cite{wadsworth1997vibrational,levine1971collision,maier1964dissociative,parks1973collision}, or by other models \cite{macheret1994mechanisms,kustova2016advanced,gonzales1993simple,knab1995theory,marrone1963chemical} in the literature. If state-specific rates are modeled with simple analytical expressions then this provides motivation to model the distributions of internal energy with analytical expressions as we propose in this article (Eq.\ref{fnb_overall}). 
The complete analysis of integrating both state-resolved rate expressions with our analytical model for $f^{NB}(j,v)$ is described in other work and is beyond the scope of the current article. However, in this section, we provide a brief description of how our nonequilibrium vibrational energy distribution model can be used to evaluate reaction rate coefficients. 

First, we have found that including overpopulation in the rotational energy distribution function has a negligible effect on the dissociation rate. However, the depletion of high-energy rotational energy states during QSS does affect the dissociation rate, compared to the assumption of a Boltzmann rotational energy distribution which contains far too many high $j$ states. Therefore, as previously published in Ref. \cite{Singhpnas}, we assume the following depleted distribution function for the rotational energy:
\begin{equation}
    f(j) \propto (2j+1) \exp\left[-\cfrac{\epsilon_{j}(j)}{{k_B T_{rot}}}  -\lambda_{1,j} \frac{\langle \epsilon_t \rangle}{\epsilon_d}  j(j+1) \right]
\end{equation}
This allows the joint rotational-vibrational distribution function, corresponding to the QSS depleted distribution (Eq. \ref{depleted_distros}) to be written as:

\begin{equation}
\begin{split}
     f^d(j,v) = \hspace{3.0in}
     \\
     \cfrac{ f^d(v;T) (2j+1)   \exp\left[-\cfrac{\epsilon_{j}(j)}{{k_B T}}  -\lambda_{1,j} \cfrac{\langle \epsilon_t \rangle}{\epsilon_d}  j(j+1) \right]}{\sum\limits_{v}^{v_{\mx}}  \sum\limits_{j}^{j_{\mx}(v)} f^d(v;T) (2j+1)  \exp\left[ -\cfrac{\epsilon_{j}(j)}{{k_B T}}  -\lambda_{1,j} \cfrac{\langle \epsilon_t \rangle}{\epsilon_d}  j(j+1) \right]}~,
       \label{dep_distros_joint}
\end{split}
\end{equation}

where the summation of $j$ levels is specific to the allowable rotational levels for each vibrational level ($v$). Since a depletion term is also included in the distribution representing overpopulation (Eq. \ref{overpop_distros}), we include both rotational and vibrational depletion in the expression for $\tilde{f}()$, which becomes:

  \begin{equation}
 \begin{split}
       \tilde{f}(j,v;T_v,T_0,T_{rot})  = \hspace{3.0in}\\
        \cfrac{ \tilde{f}(v;T_v,T_0) (2j+1) \exp\left[ -\cfrac{\epsilon_{j}(j)}{{k_B T_{rot}}}  -\lambda_{1,j} \cfrac{\langle \epsilon_t \rangle}{\epsilon_d}  j(j+1)\right]}{ \sum\limits_{v}^{v_{\mx}}  \sum\limits_{j}^{j_{\mx}(v)}\tilde{f}(v;T_v,T_0) (2j+1) \exp\left[-\cfrac{\epsilon_{j}(j)}{{k_B T_{rot}}}  -\lambda_{1,j} \cfrac{\langle \epsilon_t \rangle}{\epsilon_d}  j(j+1)\right]} \hspace{0.5in}
       \label{overpop_distros_joint}
        \end{split}
 \end{equation}
The complete nonequilibrium joint rotational and vibrational energy distribution can then be written as,
\begin{equation}
     f^{NB}(j,v) = \cfrac{\tilde{f}(j,v;T_v,T_0,T_{rot})+\Lambda f^d(j,v;T,T_{rot},T)}{1+\Lambda } 
     \label{fjvnb_overall}
\end{equation}
Finally, using Eq.~\ref{rate_eqn} and \ref{fjvnb_overall}, the expression for the rate constant is
 \begin{equation}
     k = \cfrac{  \tilde{k}(T_v,T_0,T_{rot}) +\Lambda k^d(T,T_{rot},T)}  {1 + \Lambda } ~,
      \label{rate_eqn_full}
\end{equation}
where
\begin{equation}
\begin{split}
    \tilde{k}(T_v,T_0,T_{rot}) = \sum_v^{v_{\mx}}  \sum_j^{j_\mx(v)} k(j, v; T) \tilde{f}(j,v;T_v,T_0,T_{rot}) \hspace{1.0 in}
    \label{rate_eqn_tilde}
\end{split}
\end{equation}
\begin{equation}
\begin{split}
    k^d(T,T_{rot},T) = \sum_v^{v_{\mx}}  \sum_j^{j_\mx(v)} k(j, v; T) f^d(j,v;T,T_{rot},T) \hspace{1.0 in}
    \label{rate_eqn_d_joint}
\end{split}
\end{equation}

Equations ~\ref{rate_eqn_full}--\ref{rate_eqn_d_joint} combined with Eqs.~\ref{dep_distros_joint}--\ref{overpop_distros_joint} are general and can be easily evaluated numerically provided the variation of diatomic energies ($\epsilon(j,v)$) against quantum number is known, which in the current work are based on \textit{ab initio} PESs\cite{paukku2013global,paukku2014erratum, paukku2017potential}. However, for computational cost savings in CFD solvers, numerical evaluation of the sums in Eqs.~\ref{rate_eqn_full}--\ref{rate_eqn_d_joint} can be replaced by analytical sums using simplified approximations to $\epsilon(j,v)$. Separation of energies into vibrational and rotational quantum states can also be carried out using a vibration prioritized framework \cite{jaffe1987} ($\epsilon_v(v) =\epsilon(0,v)$ and $\epsilon_j(j) = \epsilon(j,v) - \epsilon(0,v)$). Rotational energy and vibrational energy can then be fit to rigid rotor and harmonic oscillator (or some variants of rigid rotor and harmonic oscillator). These distributions can be used to obtain the non-equilibrium correction factors numerically or analytically (the ratio of non-Boltzmann distribution based rates to the corresponding Boltzmann rates using different state-specific models). 


\section{Evaluating Recombination Rates}
Rapid expansion flows are typically characterized by higher vibrational temperature relative to translational temperature. In terms of molecular interactions, an overall balance between vibrational energy de-excitation due to trans-vibrational energy exchange, vibrational-vibrational (V-V) energy exchange, exchange reactions, and  dissociation/recombination reactions will dictate the molecular internal energy populations. While many of these processes have been studied using ab-intio approaches, such as QCT and DMS, we do not have ab-intio results for recombination reactions. It is important to note that the nonequilibrium vibrational energy distribution model proposed in this article is constructed based on ab-intio data for internal energy excitation and dissociation. Currently recombination processes are not simulated in the DMS calculations used to develop the model. 

In absence of such data, we recommend to only use the overpopulation contribution of the new model when the local translational temperature is greater than or equal to the vibrational temperature. When the opposite is true, only the depletion term should be used (i.e. $\Lambda^{-1} = 0$, for $T < T_v$):
 \begin{equation}
     k = \cfrac{  \tilde{k}(T_v,T_0,T_{rot}) +\Lambda k^d(T,T_{rot},T)}  {1 + \Lambda } ~, T \ge T_v
\end{equation}
 \begin{equation}
     k = k^d(T,T_{rot},T) ~, T < T_v
\end{equation}

The recombination rate in any region of the flow can then be calculated from the dissociation rate using detailed balance ($k_{EQ}$ is equilibrium rate constant). 
 \begin{equation}
     k_{rec} = k^d(T,T_{rot},T)/k_{EQ} ~,
\end{equation}

In rapid expansions where $T_v$ maybe frozen significantly higher than T, the above expression would not include overpopulation of the distribution function and any associated coupling with the dissociation and recombination rates. However, these will be second-order effects, since the overall rates are still controlled by the forward dissociation rate and detailed balance. We note that by assuming mono-quantum transitions dominated by vibration-vibration (v-v) exchanges, the non-Boltzmann distributions have been modeled analytically by Treanor \textit{et al.} \cite{treanor1968vibrational}. However, the conditions for which the model is valid are outside of the shock-dominated or high temperature flows. For instance, whether the high lying quantum states will adjust to low translational temperature resulting in depletion, or overpopulation due to v-v exchanges and recombination, require \textit{ab initio} simulation data. The modeling of expanding flows therefore will require an independent investigation using DMS with recombination reactions, which is beyond the scope of the current work. Therefore the current model is aimed at improving the prediction of dissociating flows while reverting to the current standard model in regions where recombination is important.

\section{Conclusions}
We propose a simple generalized  model for non-Boltzmann vibrational distributions for gases in non-equilibrium, relevant for high-temperature conditions. This has been possible due to capability to perform DMS of non-equilibrium gas systems which require \textit{PESs} as the only input.   We have proposed a simple model, where the overpopulation during rapid ro-vibrational excitation is approximated by bi-modal distribution at $T_v$ and $T$ and the depletion due to dissociation of the gas in the steady state is modeled using the surprisal framework developed in the earlier work \cite{Singhpnas}. The model is demonstrated to capture the non-equilibrium distributions in the transient and QSS phase observed in DMS simulations for ro-vibrational excitation of oxygen and nitrogen systems for constant temperature and constant energy relaxations. With the proposed accurate representation of the non-Boltzmann distributions, the state-specific reaction (mainly dissociation) can be integrated over these distributions (instead of integrating over equilibrium Boltzmann
distributions) to obtain the non-equilibrium rate constants. In this manner, an analytical continuum level rate expression resembles the master-equation evolution, avoiding the computational cost involved in the master equation. 

\section*{Acknowledgement}
This work was supported by Air Force Office of Sci-
entific Research Grants  FA9550-16-1-0161 and  FA9550-19-1-0219 and was also partially supported
by Air Force Research Laboratory Grant FA9453-17-2-0081. Narendra Singh was partially supported by Doctoral Dissertation Fellowship at the University of Minnesota.

\bibliography{Bib_CFD_DSMC}

\end{document}